\documentclass[british]{INTERSPEECH2023_arxiv}

\usepackage[utf8]{inputenc}
\usepackage[T1]{fontenc}
\usepackage[british]{babel}

\usepackage{pifont}

\usepackage{graphicx}
\usepackage{epsfig}
\usepackage{adjustbox} 
\usepackage{xcolor}
\graphicspath{{./figures/}}
\DeclareGraphicsExtensions{.pdf,.png,.jpeg,.jpg}

\usepackage{amsmath}
\usepackage{amssymb}
\usepackage{bm} 
\usepackage{esint} 
\usepackage{commath} 

\usepackage{float} 
\usepackage{dblfloatfix} 
\usepackage{subcaption} 

\usepackage{booktabs} 
\usepackage{array} 

\usepackage{multirow} 
\usepackage{rotating} 

\usepackage{enumitem} 
\setlist{nolistsep} 

\usepackage[unicode=true, pdfusetitle,%
pdfauthor={Pablo Pérez Zarazaga, Zofia Malisz, Gustav Eje Henter, Lauri Juvela},%
pdfkeywords={speech synthesis, formant synthesis, neural vocoding},%
bookmarks=true, bookmarksnumbered=true, bookmarksopen=false,%
breaklinks=true, pdfborder={0,0,0}, colorlinks=true, citecolor=blue]{hyperref}

\usepackage[capitalize]{cleveref} 
\Crefname{section}{Section}{Sections}
\crefname{section}{Sec.}{Secs.}
\Crefname{table}{Table}{Tables}
\crefname{table}{Tab.}{Tabs.}

\renewcommand{\mid}{\,\ifnum\currentgrouptype=16 \middle\fi|\,}

\let\oldmarginpar\marginpar
\renewcommand\marginpar[1]{\-\oldmarginpar[\raggedleft\footnotesize #1]%
{\raggedright\footnotesize #1}}

\interspeechcameraready

\title{Speaker-independent neural formant synthesis}
\name{Pablo Pérez Zarazaga$^1$, Zofia Malisz$^1$, Gustav Eje Henter$^1$, Lauri Juvela$^2$ \thanks{This work was supported by the Swedish Research Council grant no.\ 2017-02861 ``Multimodal encoding of prosodic prominence in voiced and whispered speech'' and partially supported by the Wallenberg AI, Autonomous Systems and Software Program (WASP) funded by the Knut and Alice Wallenberg Foundation. We acknowledge the computational resources provided by the Aalto Science-IT project.}}
\address{%
$^1$Division of Speech, Music and Hearing, KTH Royal Institute of Technology, Sweden\\
$^2$Department of Information and Communications Engineering, Aalto University, Finland%
}
\email{\{\href{mailto:pablopz@kth.se}{pablopz}, \href{mailto:malisz@kth.se}{malisz}, \href{mailto:ghe@kth.se}{ghe}\}\href{mailto:ghe@kth.se}{@kth.se}, \href{mailto:lauri.juvela@aalto.fi}{lauri.juvela@aalto.fi}}

\begin{document}

\pagestyle{plain}

\maketitle
 
\begin{abstract}
We describe speaker-independent speech synthesis driven by a small set of phonetically meaningful speech parameters such as formant frequencies. The intention is to leverage deep-learning advances to provide a highly realistic signal generator that includes control affordances required for stimulus creation in the speech sciences. Our approach turns input speech parameters into predicted mel-spectrograms, which are rendered into waveforms by a pre-trained neural vocoder. Experiments with WaveNet and HiFi-GAN confirm that the method achieves our goals of accurate control over speech parameters combined with high perceptual audio quality. We also find that the small set of phonetically relevant speech parameters we use is sufficient to allow for speaker-independent synthesis (a.k.a.\ universal vocoding).
\end{abstract}
\noindent\textbf{Index Terms}: speech synthesis, formant synthesis, neural vocoding

\section{Introduction}
\label{sec:intro}
The quality of synthetic speech audio has advanced dramatically in the last decade, and can now rival recorded natural speech in terms of realism \cite{malisz2019modern}.
This development has largely come to pass by replacing methods based on signal processing with data-driven approaches such as neural vocoders \cite{tamamori2017speaker,kong2020hifi}.

Not all applications of speech technology have benefited from these advances, however.
A case in point is speech science, e.g., phonetics or psycholinguistics. In these disciplines, synthetic stimuli are used in e.g., experiments that disentangle perceptual effects of different acoustic speech properties.
To build the stimuli, speech parameters such as pitch and formant frequencies, need to be carefully controlled \cite{van1999categorical, lisker1970voicing}.
Unfortunately, contemporary signal generators based on deep learning do not offer the ability to accurately control such low-level signal aspects.
As a result, phonetics and other speech sciences remain reliant on legacy tools such as formant synthesis \cite{klatt1990analysis,carlson1982multi} or overlap-add techniques \cite{charpentier1986diphone,dutoit1996mbrola} for stimulus creation.
However, speech from such systems is not convincingly realistic: it lacks many of the cues present in natural speech and is perceived differently by humans as evidenced in slower cognitive processing times \cite{winters2004perception, van1999categorical}.
This casts some doubts over research findings based on stimuli generated by such legacy tools \cite{winters2004perception}.


This paper aims to show a way in which advances in neural vocoding can benefit speech scientists by providing speech signals that are both realistic and controllable.
Concretely, we describe a system that predicts acoustic features (mel-spectrograms) for waveform synthesis from phonetically meaningful speech parameters.
We use the term \emph{neural formant synthesis} for the idea of mapping of speech parameters to waveforms using deep learning.
Further goals are: 1) speaker-independence, to synthesise speech in any voice, and 2) use of pre-trained systems,
to make the approach easy to adopt.
We validate the proposed approach in objective experiments on control accuracy, as well as through subjective listening tests comparing to conventional neural vocoders\footnote{Audio examples and links to the code and models can be found here: \url{https://perezpoz.github.io/neuralformants}}.
These experiments show that speaker-independent neural formant synthesis leveraging HiFi-GAN reaches better subjective quality and more accurate speech-parameter reproduction than a WaveNet vocoder driven by conventional mel-spectrograms.

\section{Related work}
\label{sec:background}

\subsection{Neural vocoders}
Beginning with WaveNet \cite{vandenoord2016wavenet,tamamori2017speaker}, recent years have seen an explosion of \emph{neural vocoders} - speech signal generators based on neural networks.
These have substantially advanced the realism of synthetic speech signals.
Virtually every generative deep-learning paradigm has been investigated for neural vocoding:
autoregressive density models using dilated CNNs \cite{vandenoord2016wavenet} or RNNs \cite{mehri2017samplernn,kalchbrenner2018efficient,valin2019lpcnet}, GANs \cite{bollepalli2017generative,juvela2019gelp,kong2020hifi}, normalising flows \cite{prenger2019waveglow,kim2019flowavenet,wang2019neural}, and denoising diffusion models \cite{kong2021diffwave,chen2021wavegrad}.
Although neural vocoders in principle can be trained to generate waveforms from any audio representation, the most common choice
is to generate output from
mel-spectrogram acoustic features, e.g., as in \cite{shen2018natural,prenger2019waveglow,kong2020hifi}.
These features have proven useful for text-to-speech systems, but they do not provide an intuitive way to control important phonetic properties of speech.
There are a few neural vocoders, e.g., \cite{bollepalli2017generative,valin2019lpcnet,juvela2019gelp,wang2019neural}, that leverage the source-filter model of speech production \cite{fant1960acoustic}.
These offer a better starting point for pitch manipulation, but not necessarily other speech properties such as formant frequencies.

\begin{figure*}[!t]
    \centering
    \includegraphics[width=0.99\textwidth]{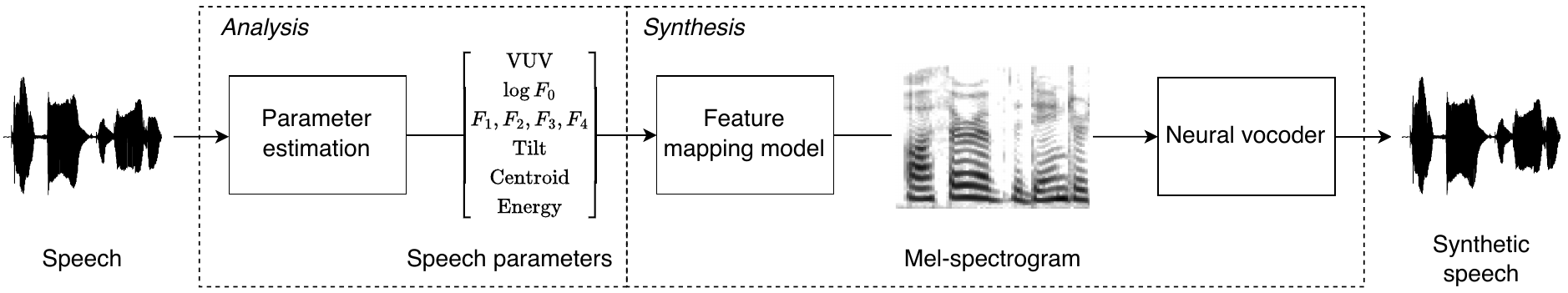}
    \caption{Block diagram of our neural formant synthesis approach. 
    %
    At analysis stage, we estimate a compact set of speech parameters, consisting most notably of the fundamental and formant frequencies. The synthesis stage consists of mapping these compact parameters into mel spectrograms that in turn drive a generic neural vocoder for waveform synthesis.
    }
    \label{fig:system-block-diagram}
    \vspace{-1\baselineskip}
\end{figure*}

\subsection{Speech parameter manipulation}
As existing neural vocoders provide little or no ability to control meaningful properties of the speech signal, they have remained largely unused in speech and language sciences, which still depend on older approaches instead.
The tools used relate to the speech aspect that needs to be manipulated in a given linguistic experiment.
Pitch and timing manipulation is probably implemented best e.g., using tools based on pitch-synchronous overlap-add (PSOLA) and its extensions \cite{charpentier1986diphone,dutoit1996mbrola}, or via signal-processing-based vocoders such as STRAIGHT \cite{kawahara2006straight}.
Manipulating other speech properties, such as formant frequencies, is more involved and difficult to achieve without degraded speech quality.
A widely used tool for formant manipulation is Praat \cite{praat_2023}, which combines linear predictive analysis with root-finding to translate the LPC polynomial into formant frequencies and bandwidths. While minor formant modification still allows the use of the original excitation source signal, more drastic modification can create artefacts due to imperfect source-filter separation, resulting in clashing residue formant information in the source signal. For larger modification, Praat can create a resonance-free source signal using the classic impulse train method, but this simplified approach degrades the quality of the synthesis. 
For optimal speech signal control, one might use a conventional formant synthesiser to generate synthetic speech waveforms with formant trajectories etc.\ completely determined by the experimenter.
However, even though formant synthesis is capable of high signal quality \cite{hunt1986generation}, the results achieved when using these systems in practice leave a lot to be desired in terms of speech realism \cite{winters2004perception,malisz2019modern}.

The most similar prior work is probably Wavebender GAN \cite{beck2022wavebender}, 
a model able to generate high-quality synthetic speech from a small set of phonetically relevant core speech parameters using a pre-trained neural vocoder.
We use a closely related set of speech parameters, but whereas Wavebender GAN is single speaker and requires several networks to be trained, we present a simpler architecture. We show that it successfully generates speech also with a variety of unseen voices.
We also study the effect of using different neural vocoders with the approach.

\section{Method}
\label{sec:method}
The goal of this paper is to generate speech signals from phonetically meaningful speech parameters, as in classic formant synthesis, using neural waveform generators for better speech quality.
To do this, we use multispeaker data to train a small network that converts speech-parameter trajectories as input into mel-spectrograms, which then drive a pre-trained neural vocoder.
See Fig.\ \ref{fig:system-block-diagram} for an overview.
In this section, we present the speech parameters to be controlled and the approach to speaker independence,
as well as detail the network used to predict mel-spectrogram acoustic features from speech parameters.



\subsection{Speech parameter extraction}
\label{sec:features}
For the controllable neural formant synthesis in this paper, we include a core set of phonetically meaningful speech parameters.
Specifically, we decided to stay close to the parameters selected for the demonstration system in Wavebender GAN \cite{beck2022wavebender}.
The speech parameters we include are:
\begin{itemize}
\item A voice flag: A binary flag indicating if the current frame is voiced or unvoiced (VUV).
\item Log $F_0$: log-scale fundamental frequency of the signal. 
\item $F_1$--$F_4$: Values of the first four formants as resonant frequencies of the vocal tract on the linear scale.
\item Spectral tilt: Rate of decay of the spectrum's energy towards higher frequencies, measured as the slope of the linear regression of all the points in the magnitude spectrum of a frame.
\item Spectral centroid: This represents the ``centre of gravity'' of the linear magnitude spectrum, computed as an average of the frequencies of all frequency-bin centre frequencies, weighted by the fraction of the total spectral energy in each bin.
\item Energy: Signal energy of the frame in linear scale (not used in Wavebender GAN).
\end{itemize}
The parameters are extracted for every analysed frame, using an analysis window of $0.0464$~s and a hop size of $0.0116$~s, which correspond to 1024 and 256 samples at the 22,050\,Hz sampling rate. The parameters make a total of one binary flag and eight continuous-valued features per each analysis frame. Each parameter trajectory except the binary flag was mean and variance normalised prior to applying deep learning.

$F_0$ is extracted using the algorithm implemented in Praat with a $10$~ms time step, a range of $75$--$300$\,Hz for male speakers and $100$--$500$\,Hz for females. Formant values are extracted using the Burg method, with $25$\,ms window size, and pre-emphasis over $50$\,Hz. The maximum number of formants is set to $5$ with the formant ceiling at $5000$ and $5500$\,Hz for males and females respectively. The ceilings for $F_1$ and $F_2$ are then optimised using Escudero's method~\cite{escudero2009cross}.
Log $F_0$-values were linearly interpolated in unvoiced regions.
Similar interpolation was applied for missing formant values, which may happen when the formant extractor only identifies three or fewer resonant frequencies below the Nyquist frequency.







\subsection{Speaker independence}
Conventional neural vocoders are ``universal'', i.e., the mel-spectrogram features they use as input convey enough information about the speaker's voice to achieve good speaker similarity \cite{lorenzo2018towards}.
It is important for speech scientists to be able to generate speech stimuli in any given voice, without having to access a large speaker-specific speech database.
Wavebender GAN, as a speaker-specific system requiring large amounts of audio data from the target speaker, was unable to generalise to unseen voices in preliminary experiments.
We will show that our proposed method (trained on many speakers) is able to generalise to numerous unseen voices, despite using only the small selection of speech parameters listed in the previous section.

\subsection{Feature-mapping neural network}
The feature-mapping network takes in the frame-rate speech parameters and maps those to a sequence of mel-spectrogram acoustic features.
We use a non-causal WaveNet-style gated convolutional architecture similar to \cite{Rethage2018-wavenet-denoising, airaksinen2019-nn-f0-esimation}.
The network uses a kernel width of 3, and 1024 channels for residual, skip and post-net, with a total of 6 residual blocks set to follow the dilation pattern (1, 2, 4, 1, 2, 4). We train the network on the VCTK training set~\cite{yamagishi2019vctk} for 99k updates using the Adam optimiser with learning rate 1e-4 on batches of 128 sequences 46 frames long, and using mean squared error (MSE) as the loss function.

\subsection{Vocoders}
Once acoustic features have been predicted by the feature-mapping network, these are passed to a neural vocoder for audio waveform generation.
We compare two different vocoders for this, namely WaveNet \cite{vandenoord2016wavenet,tamamori2017speaker} (autoregressive) and HiFi-GAN \cite{kong2020hifi} (non-autoregressive, GAN-based).
Both of these represent stronger, more realistic-sounding vocoders than the Griffin-Lim system considered in \cite{malisz2019modern}.
Therein, the Griffin-Lim system was found not to lead to slower cognitive processing times than in natural speech. It can, therefore, be argued that even higher synthesis quality should result in similarly low cognitive load.

The two vocoders operate at 22,050\,Hz and were trained to synthesise waveforms from the same standardised acoustic representations, namely those extracted by the Nvidia implementation of Tacotron 2 \cite{shen2018natural}.
While HiFi-GAN used the pre-trained V1 universal model, we trained a 30-layer WaveNet on the training portion of the dataset. The model uses 64 residual and skip channels, a kernel width of 3, and the dilation pattern grows in powers of two, reset every 10 layers. The model was trained for 1M updates with an autoregressive Gaussian density model \cite{ping2018clarinet}. We used the Adam optimiser at a fixed learning rate of 1e-3
and batches of 8 one-second audio segments.

\section{Evaluation}
\label{sec:experiments}
In order to assess both the accuracy of speech parameter control as well as the quality (realism) produced by our approach, we evaluate neural formant synthesis with two different neural vocoder back-ends.
We conduct copy synthesis experiments as well as tests of manipulated speech.
The following sections describe the experiments and their results in detail.

\begin{figure}[!t]
\centering
\includegraphics[width=.9\linewidth, trim={0 0 0 1.7cm}, clip]{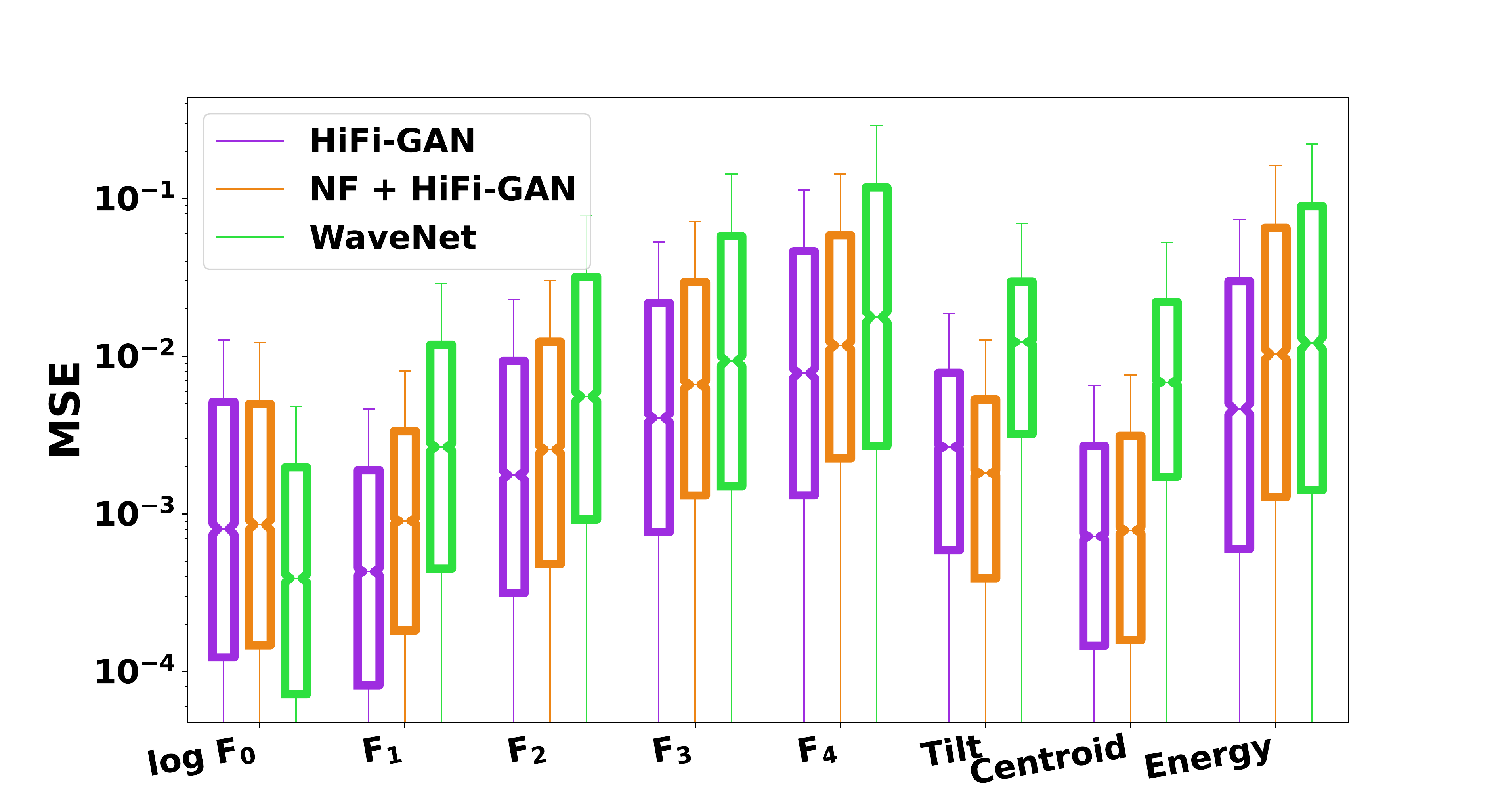}
\caption{Log-scale plot of copy synthesis results (MSE of  $z$-scored speech features). HiFi-GAN vocoding, neural formant synthesis with HiFi-GAN, and WaveNet vocoding are compared.}
\label{fig:CS_Objective}
\vspace{-1\baselineskip}
\end{figure}

The presented models were trained and evaluated on data from the VCTK corpus \cite{yamagishi2019vctk}, which contains 44,000 utterances uniformly divided among 110 English speakers with a variety of accents.
The dataset was split into training, validation, and test sets with a ratio of 80-10-10\% of the different speakers in each respective subset.
We resampled the speech to 22,050\,Hz 
and used a voice activity detector 
to trim extraneous silences in training and validation utterances.

\label{sec:systems}

In the evaluation, we compare our proposed method, i.e., feature mapping followed by a neural vocoder, to versions of WaveNet \cite{vandenoord2016wavenet,tamamori2017speaker}
and HiFi-GAN \cite{kong2020hifi} used as regular neural vocoders.
This allows to differentiate between effects due to the neural vocoder and effects due to the feature-mapping network.
Wavebender GAN, a speaker-specific system, was not included in the comparison, since we perform our evaluation on multiple speakers, none of which has enough data to feasibly train a Wavebender GAN.
However, for experiments on formant and $F_0$ manipulation, we compare to a source-filter pipeline using Praat \cite{praat_2023}
to perform the same manipulation.


\subsection{Copy synthesis evaluation}
\label{sec:objev}
\begin{figure}[!t]
\centering
\includegraphics[width=.9\linewidth, trim={0 0.2cm 0 1.7cm}, clip]{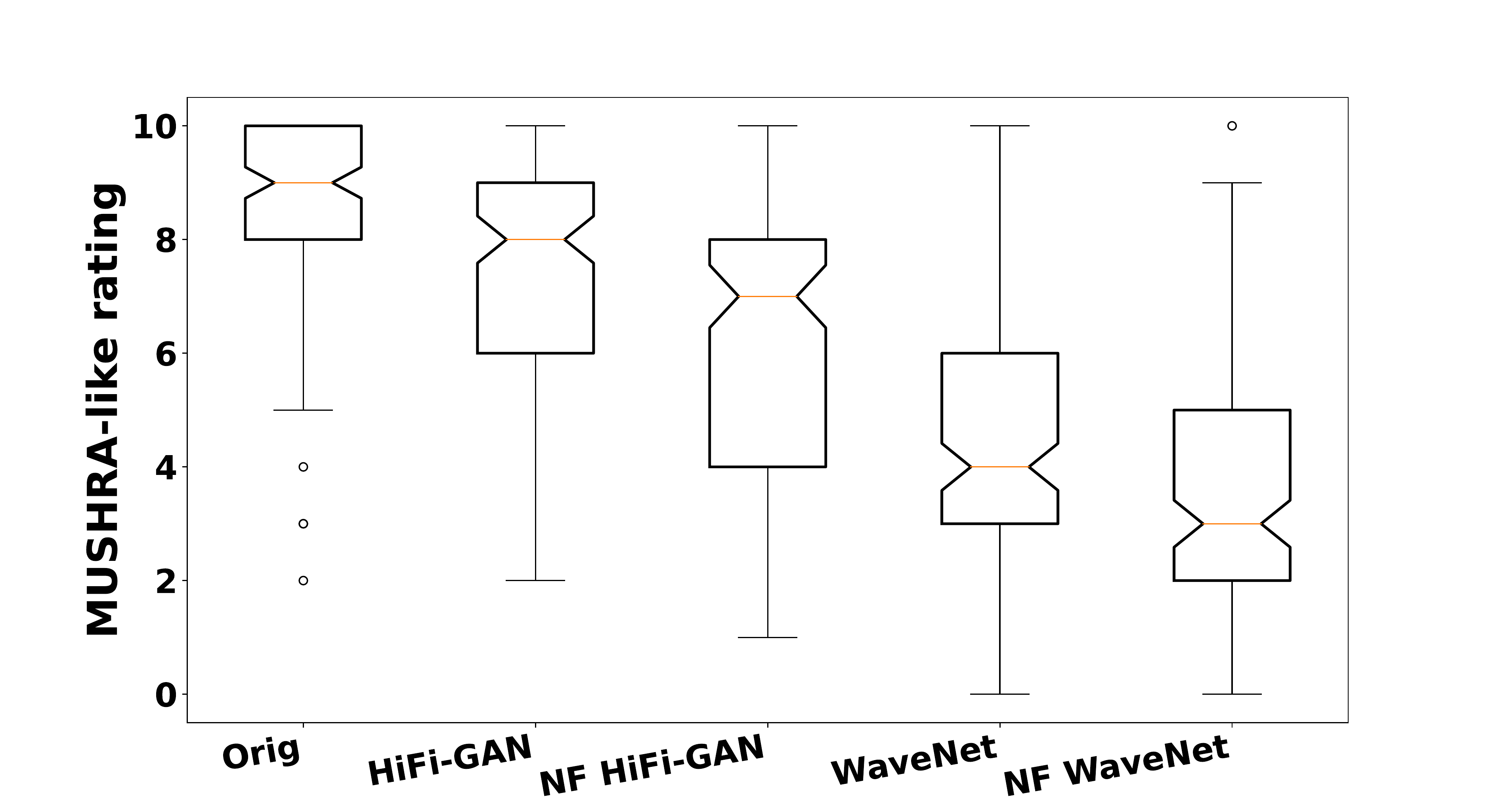}
\caption{Boxplot of listening test scores of neural vocoders and neural formant synthesis using the same two vocoders.}
\label{fig:CS_Subjective}
\vspace{-1\baselineskip}
\end{figure}
We begin by studying the accuracy of the neural formant synthesis (henceforth NF) in reconstructing speech signals using objective and subjective measurements. 
First, we perform copy synthesis with each system described in \ref{sec:systems} using speech parameters (or, for neural vocoders, acoustic features) extracted from natural speech utterances as input.
In order to compare the faithfulness of speech parameter recreation, we run speech parameter extractors on the copy synthesised audio and compare the resulting parameter trajectories to those extracted from the original speech. 
Specifically, we subtract reference values from extracted parameters in each frame, and compute the MSE across the entire test set.
We perform this error computation on $z$-score normalised values for each parameter 
quantifying the relative error for a direct comparison across speech parameters.

\begin{figure*}[!t]
\centering
\includegraphics[width=\textwidth, trim={0 0.3 0 0.7cm}, clip]{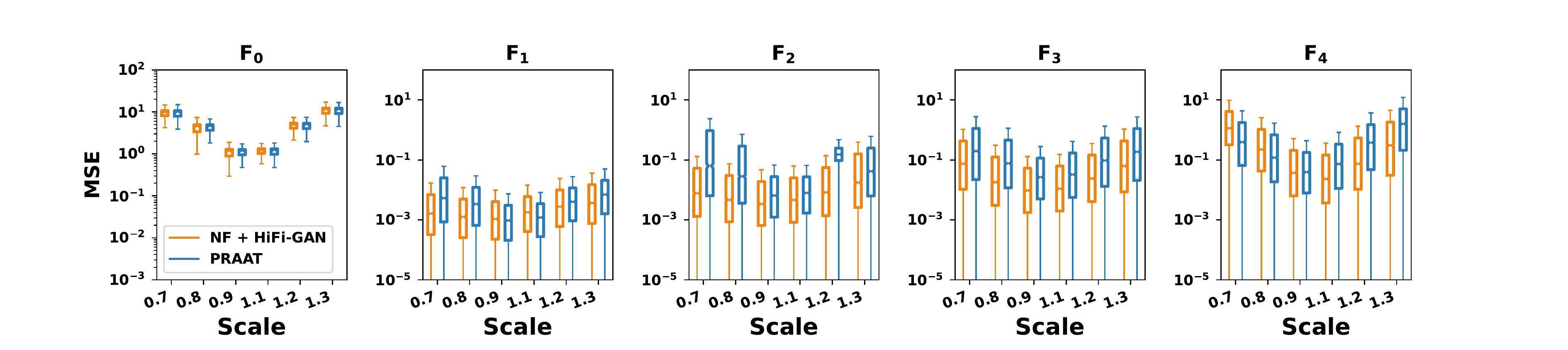}
\caption{Log-scale manipulation results (MSE of $z$-scored speech features). Each feature is scaled in the range $[0.7,..., 1.3]$ and re-synthesised using neural formant synthesis with HiFi-GAN, and then compared to using Praat for the same manipulation.}
\label{fig:manip_objective}
\vspace{-1\baselineskip}
\end{figure*}

The results of the copy synthesis experiments on utterances from held-out speakers are shown in Fig.~\ref{fig:CS_Objective}.
We see that HiFi-GAN alone best approximates natural speech parameters, compared to NF+HiFi-GAN formant synthesis.
However, the formant synthesiser is not expected to, and indeed does not, surpass HiFi-GAN's performance, with the exception of random fluctuations, as it relies on the vocoder. 
That said, the difference between HiFi-GAN and NF+HiFi-GAN is encouragingly small.
Importantly, the NF+HiFi-GAN tends to produce more faithful speech parameter trajectories than the solo WaveNet.

The median mean squared error for most speech parameters is between $10^{-4}$ and $10^{-2}$, which we consider a good result.
Log $F_0$, $F_1$, and spectral centroid achieve the lowest copy synthesis error, with errors increasing with each formant.
This trend is consistent with the formant extractor missing a resonant peak in the spectral envelope on occasion, affecting estimated formant frequencies above that peak.
$F_4$ and energy display the greatest error, despite the fact that energy is straightforwardly defined and easy to manipulate without machine learning.

To assess the subjective quality of the generated speech audio, we perform a MUSHRA-like \cite{itu2015method} test, asking crowdsourced listeners to ``Listen to the speech samples and rate the quality of the sound''. Technical limitations of the testing platform required a discrete grading scale, and in order to provide a higher resolution than a 1-5 MOS scale, we choose an integer scale from 0 (worst) through 10 (best) for their ratings.
Stimuli for the same utterance were presented side-by-side, with reference audio (the natural recorded utterance) both available as an explicit and hidden reference among the stimuli to be rated.
25 participants, all self-reported native English speakers, were recruited using the Prolific platform and ran the test in a browser, wearing headphones. 
The test included 10 utterances selected from the 10 held-out speakers. Three attention checks were used during the test. We only included copy synthesis stimuli in this evaluation, since the parameter manipulations would give rise to improbable and less natural parameter trajectories even if the manipulation is perfectly faithful.




The results of the listening test are presented in Fig.~\ref{fig:CS_Subjective}.
As expected, recorded speech achieves the highest rating, whilst neural formant stimuli each score slightly below their respective vocoder, considered an upper bound.
A key finding is that the perceived quality of NF+HiFi-GAN exceeds the quality of WaveNet driven by more information-rich mel-spectrograms.
WaveNet has established itself as the baseline high quality neural vocoder. Being able to surpass its performance with speaker-independent neural vocoding is an impressive result that indicates the strong potential our tools have for improved stimulus creation in speech and language sciences.

\subsection{Parameter-manipulation evaluation}

In the second experiment, we aim to assess the control accuracy of the present approach for manipulating the selected speech parameters listed in Sec.~\ref{sec:features}.
We study the effect of manipulating parameter trajectories fed into neural formant synthesis.
%
Specifically, manipulated speech was created by scaling each input parameter trajectory, multiplying it with a constant factor (0.7, 0.8, 0.9, 1.1, 1.2, and 1.3) across an entire utterance.
For log $F_0$, which already has been mapped to the logarithmic domain, we added or subtracted a constant value from the trajectory instead.
The scaling was performed for each utterance and for each speech parameter in isolation, keeping all other parameter trajectories as they were during the copy synthesis.
In this experiment, only
the HiFi-GAN vocoder was considered.

After having generated synthetic speech, we again ran automated speech parameter extraction on the resulting samples, to check if the parameter trajectories in the manipulated speech accurately hit their target values.
The lower and upper limits on formants and fundamental frequency for the feature extractors were adjusted as needed, so that our target would not fall outside the range considered by these feature extractors.

The results of manipulating fundamental frequency and formant frequencies are presented in Fig.~\ref{fig:manip_objective}, where they are compared to the effect of using Praat for performing the exact same manipulation.
We see that both methods have very similar accuracy when manipulating the fundamental frequency.
When manipulating formants, the neural formant synthesiser generally gives more accurate results, according to our automated analysis.
This is promising, but it is not possible to tell if these numbers are affected by the fact that errors typically made by the automated speech-parameter extractors have been seen by the neural approach during training.
In general, errors are larger, the greater the magnitude of the manipulation, both for our neural formant synthesis and the Praat baseline.

For every manipulated stimulus, we extracted the trajectories of all speech parameters to verify that parameter control is properly de-correlated and that changing one parameter does not increase the errors in another.
There is not enough space to graph these findings here, but from inspecting the output, results are mixed: neural formant synthesis exhibits larger errors in unmanipulated features than Praat when changing fundamental frequency and $F_1$, but this situation is reversed for $F_2$.
In general, neural formant synthesis outperforms Praat in terms of spectral centroid accuracy, whereas Praat is better at maintaining the correct frame energy.
To remedy that, neural formant synthesis could be made more accurate by changing the signal gain in a post-processing step. 
We invite the reader to listen to the audio samples and experiment on the source code~\footnote{Audio examples and links to the code and models can be found here: \url{https://perezpoz.github.io/neuralformants}}.



\section{Conclusions and future work}
\label{sec:conclusion}
We have demonstrated speaker-independent neural formant synthesis, recreating speakers outside the training set using two different vocoders.
Our experiments confirm that important speech parameters can be accurately controlled, and that realism is close to state-of-the-art neural vocoding despite the minimalist set of speech parameters used to drive the synthesis.

In fact, neural formant synthesis on held-out speakers with the HiFi-GAN vocoder back-end was found to provide: a) a higher perceived speech quality as well as b) a more accurate copy synthesis than a conventional WaveNet vocoder driven by mel-spectrogram acoustic features.
We think this strong result paves the way for improved synthetic and manipulated speech stimuli in speech and language sciences.
Important future work includes integration of explicit formant control through differentiable digital signal processing (DDSP) and validation of the approach in phonetic research, e.g., by recreating classic categorical perception experiments.

\ifinterspeechfinal

\bibliographystyle{IEEEtran}
\bibliography{refs}

\end{document}